\documentclass[letterpaper]{ar-1col}
\usepackage{url}
\usepackage[numbers]{natbib}
\usepackage{amsmath,amsthm,amssymb}
\usepackage{amsfonts,amssymb}
\usepackage{graphicx}
\usepackage{mathtools}
\usepackage{bm}
\usepackage{empheq}
\usepackage{lineno}
\usepackage{wrapfig}
\usepackage[]{natbib}
\usepackage{natbib}
\usepackage[dvipsnames]{xcolor}
\usepackage{physics}
\usepackage{lipsum} 
\usepackage{tikz}
\jname{Xxxx. Xxx. Xxx. Xxx.}
\jvol{AA}
\jyear{YYYY}
\doi{10.1146/((please add article doi))}

\begin{document}

\markboth{Lai et al.}{ML for Climate Physics and Simulation}

\title{Machine Learning for Climate Physics and Simulations}

\author{Ching-Yao Lai,$^1$ Pedram Hassanzadeh,$^2$ Aditi Sheshadri$^3$, Maike Sonnewald$^4$, Raffaele Ferrari$^5$, Venkatramani Balaji$^6$ 
\affil{$^1$Department of Geophysics, Stanford University, Stanford, CA 94305, USA; email: cyaolai@stanford.edu}
\affil{$^2$Department of Geophysical Sciences and Commitee on Computational and Applied Mathematics, University of Chicago, Chicago, IL 60637, USA}
\affil{$^3$Department of Earth System Sciences, Stanford University, Stanford, CA 94305, USA}
\affil{$^4$Department of Computer Science, University of California Davis, Davis, CA 95616, USA}
\affil{$^5$Department of Earth, Atmospheric, and Planetary Sciences, Massachusetts Institute of Technology, Cambridge, MA 02139, USA}
\affil{$^6$Schmidt Sciences, USA}
}
\begin{abstract}
We discuss the emerging advances and opportunities at the intersection of machine learning (ML) and climate physics, highlighting the use of ML techniques, including supervised, unsupervised, and equation discovery, to accelerate climate knowledge discoveries and simulations. We delineate two distinct yet complementary aspects: (1) ML for climate physics and (2) ML for climate simulations. While physics-free ML-based models, such as ML-based weather forecasting, have demonstrated success when data is abundant and stationary, the physics knowledge and interpretability of ML models become crucial in the small-data/non-stationary regime to ensure generalizability. Given the absence of observations, the long-term future climate falls into the small-data regime. Therefore, ML for climate physics holds a critical role in addressing the challenges of ML for climate simulations. We emphasize the need for collaboration among climate physics, ML theory, and numerical analysis to achieve reliable ML-based models for climate applications.

\end{abstract}

\maketitle

\section{Introduction} \label{intro}
\begin{figure}[b]
  \begin{center}
    \includegraphics[width=1\linewidth]{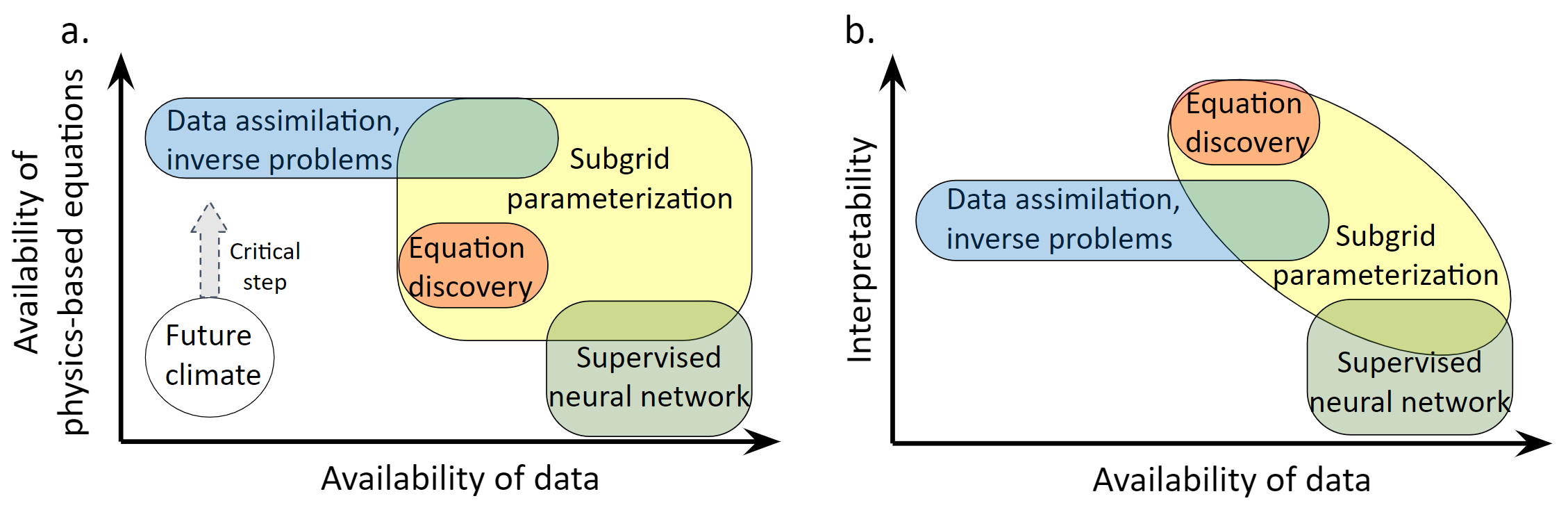}
  \end{center}
	\caption{Conceptual diagram of machine learning (ML) applications in climate sciences with respect to the availability of existing physics-based equations, availability of data and interpretability. We explain different components of this figure throughout this paper. (a) Existing physics-based equations and data are two sources of information used for training ML models. (b) More physics-based equations are not necessarily more interpretable, e.g., existing numerical weather predictions. On the other hand, equation discovery usually comes with regularization techniques to find the simplest set of equations capturing the dominant behavior of the system, enhancing interpretability. [**Note to Annual Reviews: We created this figure for this article; it is
not based on any previously published image.**]}
	\label{datamodel}
\end{figure}

\begin{figure}[t]
\centering
\includegraphics[width=0.95\linewidth]{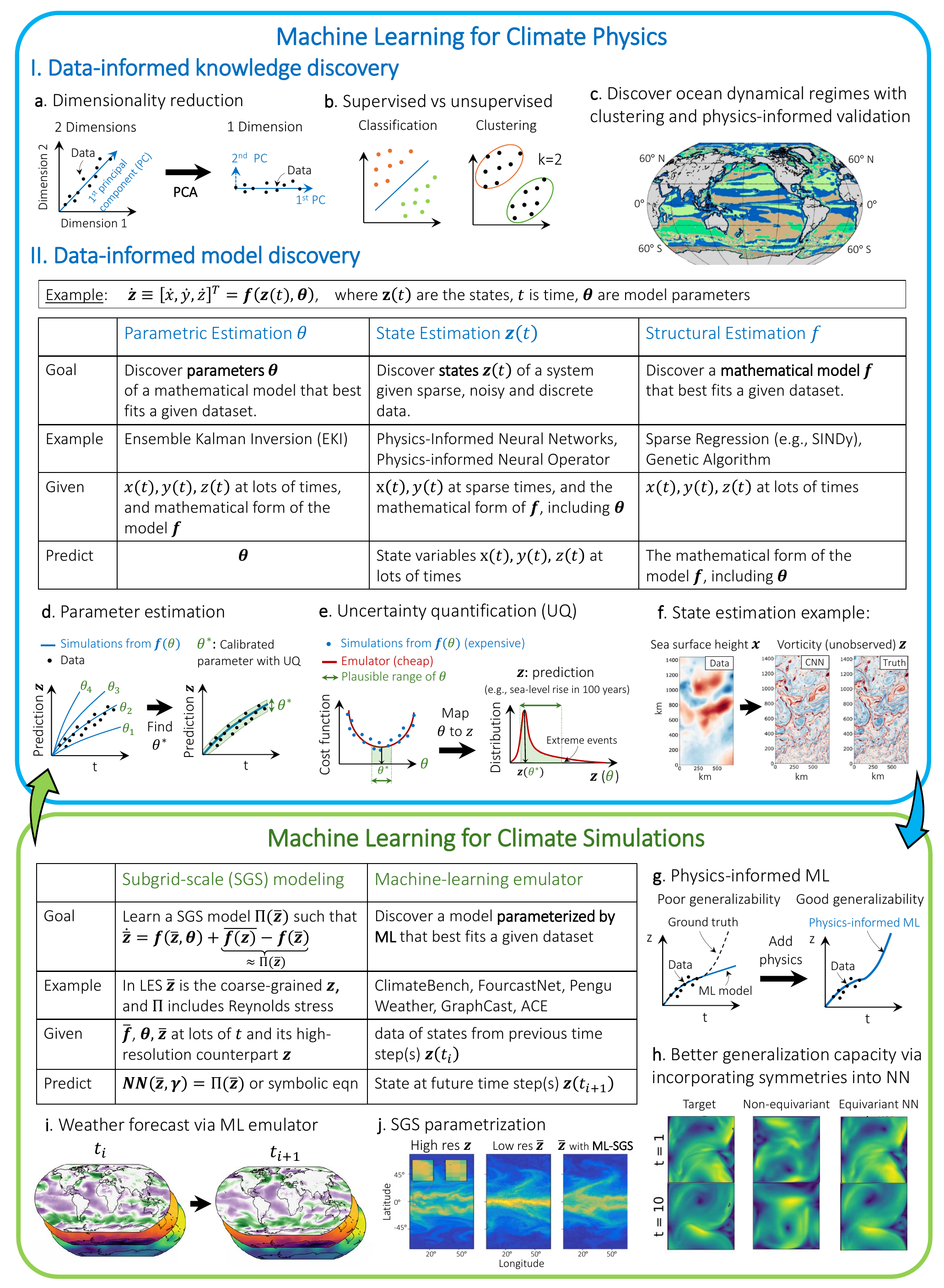} 
\caption{An overview of the areas where ML has played a role in uncovering climate physics and advancing climate simulations. Panels c, f, i, j, h adapted with permission from Sonnewald et al. 2019 \cite{sonnewald2019unsupervised} (CC BY 4.0), Xiao et al. 2023 \cite{xiao2023reconstruction} (CC BY 4.0), Lam et al. 2023 \cite{lam2023learning} (copyright 2023 AAAS), Yuval and O’Gorman 2020 \cite{yuval2020stable} (CC BY 4.0), Wang et al. 2021 \cite{wang2020incorporating} (CC BY 4.0), respectively.}
\label{fig1}
\end{figure}

Machine learning (ML) has led to breakthroughs in various areas, from playing Go to text generated with large language models (LLM), and more recently, to weather forecasting \cite{keisler2022forecasting,pathak2022fourcastnet,bi2023accurate,lam2023learning,chen2023fuxi}. Different from Go and LLM, the language scientists use to understand and simulate weather and climate has been equations rooted in fundamental physics. Physics-based equations, often differential equations, are essential to simulate systems where direct observations are limited and noisy. Even more so for projections of future climates where data are altogether unavailable. (Fig.~\ref{datamodel}). Recently, ML has emerged as an alternative tool for predictive modeling as well as improving the understanding of climate physics (Fig.~\ref{fig1}). For instance, physics-free ML models such as neural networks (NNs), which are universal approximators of functions \cite{hornik1989multilayer} and operators \cite{chen1995universal}, trained on data from observations or physics-based simulations have demonstrated remarkable ability to perform accurate nowcasting \cite{ravuri2021skilful} with lead time of a few hours, weather prediction \cite{ben2024rise} with lead times of several days, and El Nino forecast with lead times of a year \cite{ham2019deep}. It remains an open question if some of the strategies and success in these short-time predictions can be applied to improve climate projections, i.e., to estimate changes in the statistics of weather events (e.g. return periods of heat waves or tropical cyclones) in the next decades, centuries, and beyond. 

\begin{marginnote}[]
\entry{Model}{A representation of a system to make predictions. Models can be physics-based, ML-based, or coupled.}
\entry{Simulation}{Physics-based models solved numerically. For example the General Circulation Model is a climate simulation that solves the Navier-Stokes equation...etc.}
\entry{Emulator}{In this article emulator refers to a subset of models that fit the data, bypassing solving physics-based equations.}
\entry{ML-based emulator}{Emulators that fit the data with ML models (e.g., deep neural networks), bypassing solving physics-based equations.}
\entry{Neural Network (NN)}{Function approximators parameterized by function operations and parameters $\gamma$, that are optimized to minimize specified cost functions $\mathcal{L}$}
\entry{Non-stationarity}{Systems with time-evolving statistical properties, so that a limited time series is not representative of the past or the future.}
\end{marginnote}

\paragraph{Weather versus climate: non-stationarity}
The success story of ML for prediction (Section \ref{emulation}) has been primarily showcased in weather forecasting \cite{rasp2024weatherbench}. Followed by initial attempts started in 2019 (e.g., \cite{pathak2018model,dueben2018challenges,weyn2019can,chattopadhyay2020analog,rasp2020weatherbench,rasp2021data,clare2021}), by 2022-2023 ML weather models (often called ``\textit{emulators}''; Section \ref{emulation}) achieve, at a fraction of the computational cost, similar or better forecast skills than state-of-the-art physics-based weather prediction models (e.g., \cite{keisler2022forecasting,pathak2022fourcastnet,bi2023accurate,lam2023learning,chen2023fuxi,price2023gencast}). This success has generated excitement about using ML to improve climate projections as well. Yet, climate projections involve major additional challenges \cite{watson2021machine,schneider2023harnessing}. In weather forecasting, we have a constant stream of real-time and historical data for training ML-based emulators (lower-right of Fig. \ref{datamodel}a) to make predictions for a few weeks, where the statistics can be assumed to be stationary. In climate, we are often interested in predicting the climate's forced response to changes in greenhouse gases in the atmosphere, leading to ``\textit{non-stationarity}'' \cite{palmer1999nonlinear}, e.g. climate with different mean and variability. ML models are not suited to predict the behavior of a system substantially different from the one they have been trained on. Yet we simply do not have observational data for the future (i.e. lower-left corner in Fig. \ref{datamodel}a) to train and validate future predictions; this is an issue for both ML and physics-based models. We summarize the non-stationary challenge and potential solutions, such as incorporating physics constraints, in Section~\ref{sec:OOD}. Furthermore, the long-term predictions of future climate involves the interactions between the atmosphere, ocean, cryosphere, land, and biosphere, which make the problem more challenging than short-term weather forecast.

\begin{marginnote}[]
\entry{Subgrid-scale (SGS) parameterization}{A model to parametrize the relationship between the resolved states as a function of the unresolved (subgrid) state. }
\entry{ML-based parametrization}{SGS parametrization represented by a ML model rather than physics-based equations.}
\end{marginnote}

\paragraph{Challenges in understanding and simulating climate}
The climate system consists of interacting processes that span orders of magnitude in spatial (from microns to planetary) and temporal (from seconds to centuries) scales. Simulating the climate system to resolve all these scales is computationally challenging. Due to its multi-scale nature, representing physics in the under-resolved scales (e.g., cloud microphysics, turbulence) in low-resolution climate simulations—referred to as \textit{subgrid-scale (SGS) parameterization}—has been a central goal for climate scientists. ML has emerged as a promising alternative to SGS parameterization due to its ability to perform equation discovery and its desirable properties as universal function approximators, which do not require prior assumptions about the functional forms of the parameterization. We summarize recent advances of ML in SGS parameterization in Section \ref{SGS}, as well as its major challenges, such as interpretability in Section \ref{UQ} (lower-right corner of Fig. \ref{datamodel}b) and uncertainty quantification (UQ) in Section \ref{XAI}. Without understanding what the ML model actually learns and the reasoning behind it, we cannot deduce when the ML model will generalize well to future climates. One approach to improve interpretability is discovering the closed-form equations that capture the data (Section \ref{eqn}; upper side of Fig. \ref{datamodel}b). Equations have long been the language for physicists to develop understandings of the systems they govern. With the emergence of ML, ways to comprehend and extract knowledge from ML-based models are new areas of research. Several methods developed to understand what the ML model itself learns are detailed in Section \ref{XAI}.

We organize this paper focusing on two distinct goals through which ML is reshaping climate science: \textbf{(1) ML for climate physics} and \textbf{(2) ML for climate simulations}. The former focuses on utilizing the increasing availability of data from the Earth system to extract understandings, including knowledge discovery (Section \ref{discovery}) and data-driven model discovery (Section \ref{DIMD}). The latter discusses recent advances in accelerating simulations, including data-driven parameterization (Section \ref{SGS}) and climate emulators (Section \ref{emulation}). Section \ref{PIML} discusses methods used to add physical constraints to ML models. For inverse problems with sparse data (upper left of Fig. \ref{datamodel}a) in the ``small-data regime", including physical constraints are necessary to generalize prediction where data does not exist. Finally, the future climate also falls in the small-data regime, as no future observations exist. Along with an incomplete understanding of the physics of future climate, it lies at the bottom left of Fig. \ref{datamodel}a, making it the most challenging among others in Fig. \ref{datamodel}a. Thus, predicting climate is not merely about accelerating simulation but essentially requires generating more physical knowledge than currently available, moving towards the tractable upper left regime of Fig. \ref{datamodel}a. While training ML models to make accurate predictions faster than physics-based simulations, making previously challenging tasks computationally tractable, is an achievement, {\it the ability to simulate does not equate to improved physical understanding} \cite{held2005gap}. We stress that accurate and fast simulations or predictions are not sufficient; deeper physical understandings of the climate are necessary to address the climate modeling challenges.

\section{Machine Learning for Climate Physics}\label{physics}
The increase in availability of data, from both observations and high-fidelity simulations, is a key driver for new physical insights. Below we introduce two emerging trends of research using ML to improve our understanding of the climate physics using data: (1) Data-informed knowledge discovery, e.g. identifying patterns and dynamical regimes in high-dimensional, complex observations and simulations and (2) discovering data-informed predictive models.
\subsection{Data-informed knowledge discovery}\label{discovery} 
Knowledge discovery (e.g. Fig. \ref{fig1}a-c), such as identifying coherent patterns of dynamical significance in spatio-temporal data, has long been a fundamental process for making discoveries in climate science. A classical example in atmospheric science is the determination of what we now call the El Niño Southern Oscillation (ENSO) by Walker in 1928 \cite{walker1928world}. Stationed in India during the British occupation, Walker employed an army of Indian clerks to conduct principal component analysis (PCA) by hand on all available data, decomposing it into orthogonal modes that revealed coherent structures associated with ENSO. Today, we have access to vast amounts of data, both observational and computational. Advanced ML techniques emerge as powerful pattern recognition tools that computationally scale well with increasing volumes of data. Off-the-shelf tools widely adopted in climate sciences include supervised learning methods (e.g., random forest, Gaussian process regression (GPR), and NN) and unsupervised learning methods such as autoencoders and clustering algorithms (e.g., k-means, self-organizing maps). Several of the above tools have been used in climate science for decades, long before deep learning took off.

\begin{marginnote}[]
\entry{Supervised learning}{Algorithms that learn a mapping from input data (features) to output labels based on the training examples provided}
\entry{Unsupervised learning}{Algorithms that discover patterns, structures, or relationships within the input data (features) without explicit guidance from labeled examples.}
\end{marginnote}

\paragraph{Dimensionality reduction}
Climate data often involve high-dimensional, nonlinearly correlated, spatial and temporal variables, such as temperature, pressure, and precipitation, across large geographical regions and long time periods. Dimensionality reduction has long been used in for transforming high-dimensional climate data into a lower-dimensional space, which might be more amenable to physical interpretation and to develop reduced-order predictive models. 
PCA, also known as Empirical Orthogonal Function (EOF), is one of the commonly used {\it linear} techniques for dimensionality reduction (Fig. \ref{fig1}a). Reducing decades of observational data into a few modes of variability, such as ENSO, has facilitated understanding of the underlying dynamics and even the robust detection and interpretation of the anthropogenic climate-change footprint~\cite{corti1999signature,thompson2002interpretation}. However, traditional techniques such as PCA/EOF have major limitations, such as the lack of a dynamical meaning of the discovered ``modes" and the linearity \cite{monahan2009empirical}. (See Section ``Beyond EOFs: Principal Oscillation Patterns and ML" in the Supplemental Material). 

ML techniques have the potential to address the linearity limitation. For example, autoencoders, a type of NN with high-dimensional input and output layers and a lower-dimension latent space, are powerful tools for dimensionality reduction. A single-layer autoencoder with linear activation functions is mathematically equivalent to PCA. In contrast, deep autoencoders with nonlinear activation functions have more expressive power in capturing the low-dimensional representation of the high-dimensional input \cite{page2021revealing, lusch}, and their applications to climate science have started to emerge. Shamekh et al. \cite{shamekh2023implicit}, interpreting the latent space of an autoencoder, developed of a new metric for cloud and precipitation organization, enabling the development of a better parameterization for moist convection.

Markov models that describe the transition probability from one state to another are being applied to study the evolution of the climate system in a reduced space, possibly based on a PCA/EOF projection~\cite{Souza23,Geogdzayev24}. Latent variable models have also shown promise in numerical model analysis and predictive skill. For example, Wang et al. \cite{Wang2020} improved ENSO prediction skill by using kernel analog forecasting (related to the Koopman operators \cite{rowley2009spectral}).

\paragraph{Finding patterns in climate data}
The task of finding patterns in climate data extends far beyond dimensionality reduction and is a fruitful area that still has much to be explored. For example, unsupervised methods like clustering (Fig. \ref{fig1}b) have been used to identify the balance between terms of the equations governing simulation data and to discover global ocean dynamical regions as parsimonious representations of the governing equations \cite{sonnewald2019unsupervised} (Fig. \ref{fig1}c).

Data from actual observations can often inspire new knowledge about climate systems \cite{reichstein2019deep}. Supervised methods facilitate the utilization of vast amounts of satellite observations, such as reconstructing a pan-Arctic dataset of sea-ice thickness during periods when data are unavailable \cite{landy2022year}, revealing the strong nonlinear interactions of ocean eddies \cite{martin2024deep}, reconstructing ocean surface kinematics with sea-surface height measurements \cite{xiao2023reconstruction}, and detecting icebergs to understand their contribution to the freshwater budget \cite{rezvanbehbahani2020significant}. Extracting information from remote-sensing data can fill missing gaps required to inform physics-based models. For example, the identification of ice fractures (under-resolved in simulations) is needed to constrain parameters for modeling ice dynamics \cite{surawy2023episodic, lai2020vulnerability}. Some of the above tasks have long been done manually and often ``subjectively" by scientists. ML offers an efficient alternative that can be easily scaled up to all available data that may be intractable otherwise, and can be made easily accessible, reproducible, and transparent via open-source software. That said, the design of the loss functions is still subjective.

\subsection{Data-informed model discovery} 
\label{DIMD}
Apart from distilling knowledge from data, physicists have been developing predictive models to describe observations for centuries. The utility of a model, if it accurately represents the observations, lies in its ability to make predictions when the data is unavailable, such as projections about the future. The crux of climate physics is creating trustworthy future predictions, and determining how to construct a model that faithfully describes the data is essential. It's worth noting that in traditional physical sciences, a ``model" often takes the form of mathematical equations. In modern ML literature, a model can refer to functional operations (e.g., NNs) that parameterize relationships between specified input and output variables. For clarity, in this section by ``model discovery" we meant the discovery of mathematical equations. 

Here, we broadly classify three different ML approaches that have been used for finding models to describe climate data: parametric, state, and structural estimations (Fig. \ref{fig1}). We use a dynamical-system example of the following form to illustrate the differences: 
\begin{equation}
\frac{d}{dt}\bm{z}(t) = \bm{f}(\bm{z}(t), \bm{\theta}),
\label{dynamics}
\end{equation}
where $\bm{z}(t)\in \mathbb{R}^n$ represents the states vector of the system that evolves with time $t\in \mathbb{R}$, and its evolution is dictated by the mathematical expression of the dynamics $\bm{f}$ and the model parameters $\bm{\theta}$. See sidebar titled The Lorenz 63 system for an example.
\begin{textbox}[h]\section{The Lorenz 63 system}
The Lorenz 63 system \cite{lorenz1963deterministic}, a simplified mathematical model for atmospheric convection, is described by the following set of ordinary differential equations: 
\begin{equation}
\frac{d}{dt}{x}=a(y-x), \quad 
\frac{d}{dt}{y}=x(b-z)-y, \quad
\frac{d}{dt}{z}=xy-cz,
\label{lorenz}
\end{equation}
where the model parameters $\bm{\theta} = [a,b,c]$ are constants, and $\bm{z}(t)=[x(t),y(t),z(t)]$ are the time-evolving states. 
\end{textbox}
 
\paragraph{Parametric estimation $\bm{\theta}$} 
Given a discrete dataset of states measured at discrete times $t_i$, i.e. $\{x(t_i),y(t_i),z(t_i)\}_{i=1}^N$, parametric estimation refers to predicting the free parameters $\bm{\theta} = [a,b,c]$ when the functional form of the mathematical model $\bm{f}(\bm{z}(t), \bm{\theta})$ is known (Fig. \ref{fig1}d). 
\paragraph{State estimation $\bm{z}$}
State estimation involves predicting the state variables $\bm{z}(t)$ given the mathematical model $\bm{f}(\bm{z}(t), \bm{\theta})$, model parameters $\bm{\theta}$ and data $\{x(t_i),y(t_i),z(t_i)\}_{i=1}^N$. Estimating states is particularly useful for data interpolation when the available data is sparse in time or space, for data denoising, or for inversion when the predicted state is not measurable (e.g. Fig. \ref{fig1}f) and thus completely unavailable in the data library, e.g., predicting $z(t)$ with data of $\{x(t_i),y(t_i)\}_{i=1}^N$.

\paragraph{Structural estimation $\bm{f}$} Also referred to as equation discovery. Prediction of the complete mathematical expression of $\bm{f}(\bm{z}(t), \bm{\theta})$ (e.g. equation \eqref{lorenz}), including the free parameters $\bm{\theta}=[a,b,c]$ given only the discrete data $\{x(t_i),y(t_i),z(t_i)\}_{i=1}^N$. The determination of the model $\bm{f}(\bm{z}(t), \bm{\theta})$ fully relies on the data; therefore, dense data is often needed to guarantee the success of the algorithms.\\

Beyond these three categories, an emerging data-driven approach involves replacing $\bm{z}(t)$ with a ML-based emulator. In this approach, instead of using equations, the dynamics $\bm{f}$ are represented by black-box ML models, as detailed in Section \ref{emulation}. 

\subsubsection{Algorithms and examples} \label{sec:algo}
In this section, we list a few examples that demonstrate how ML has influenced data-driven model discovery within the three categories described above; ensemble Kalman inversion (EKI) for parameter $\bm{\theta}$ estimation, physics-informed ML for state $\bm{z}$ estimation, and equation discovery for structural $\bm{f}(\bm{z}, \bm{\theta})$ estimation. 

\begin{textbox}[h]\section{Inverse problem and data assimilation}
The problems of estimating parameters $\bm{\theta}$ and states $\bm{z}$ of a model $\bm{f}(\bm{z},\bm{\theta})$ falls under the umbrella of inverse problems \cite{tarantola2005inverse}. The importance of parametric and state estimations lies in the fact that direct observations of model parameters and states are often unavailable, yet they are crucial for simulating and predicting both the weather and climate accurately. Various inverse problems arise in weather and climate, such as estimating parameters in climate models or determining initial conditions for improved weather forecasting. Data assimilation \cite{kalnay2003atmospheric} refers to the process of combining observational data with numerical models.

The synergies between data assimilation and ML has been increasingly recognized \cite{geer2021learning,brajard2021combining,cheng2023machine}, including using ML to correct model error in data assimilation \cite{farchi2021using} and emulation of a dynamical system \cite{brajard2020combining}.
\end{textbox}

\paragraph{Ensemble Kalman inversion (EKI)}
Here we focus on one family of methods, EKI, as an example, as it is increasingly used in climate science. EKI \cite{iglesias2013ensemble} is a well-developed parameter estimation technique in the climate modeling community and has been used in various contexts such as convection, turbulence, and clouds \cite{cleary2021calibrate,dunbar2021calibration,lopez2022training}, gravity waves \cite{mansfield2022calibration}, and ocean convection \cite{souza2020uncertainty}. EKI is a derivative-free \cite{iglesias2013ensemble} optimization method for parametric estimation $\theta$, based on Ensemble Kalman filtering (EnKF) \cite{evensen1994sequential}, which is used for estimating states $\bm{z}(t)$ in numerical weather prediction (NWP) \cite{houtekamer2016review} given noisy observations. EKI \cite{kovachki2019ensemble} attempts to find a distribution of model parameters $\bm{\theta}$ that can describe time-averaged statistics of a ``truth", which could be from observational data or simulations, removing dependence on state variables by utilizing long integrations. EKI optimizes for macrophysical climate statistics (e.g., derived by averaging over many occurrences of the event of interest).  

Quantification of parametric uncertainty is important as it illustrates how perturbations of parameters $\theta$ that we want to estimate would translate to the predictions of $z$. As shown in Fig. \ref{fig1}e, while an optimal value of $\theta^*$ minimizing the cost function only captures one prediction $z(\theta^*)$, a range of $\theta$ could yield wide-ranging predictions (e.g., covering extreme events). Running ensembles of forward physics-based simulations $\bm f(\theta)$ with a range of $\theta$ for climate models to quantify these uncertainties propagated by parameter uncertainty is currently computationally infeasible. To address this challenge, the calibrate-emulate-sample (CES) approach \cite{cleary2021calibrate,dunbar2021calibration} trained a Gaussian process regression (GPR) as a cheap method to emulate the prediction of interest $z$ as a function of $\theta$. Sampling the GPR emulator with Markov chain Monte Carlo enables substantially faster uncertainty quantification of the predictions $z$ resulting from the plausible range of $\theta$. GPR has also been used directly for calibrating parameters with uncertainty quantification (UQ) in Earth system models \cite{watson2021model}.

\begin{figure}[h]
\centering
\includegraphics[width=1\linewidth]{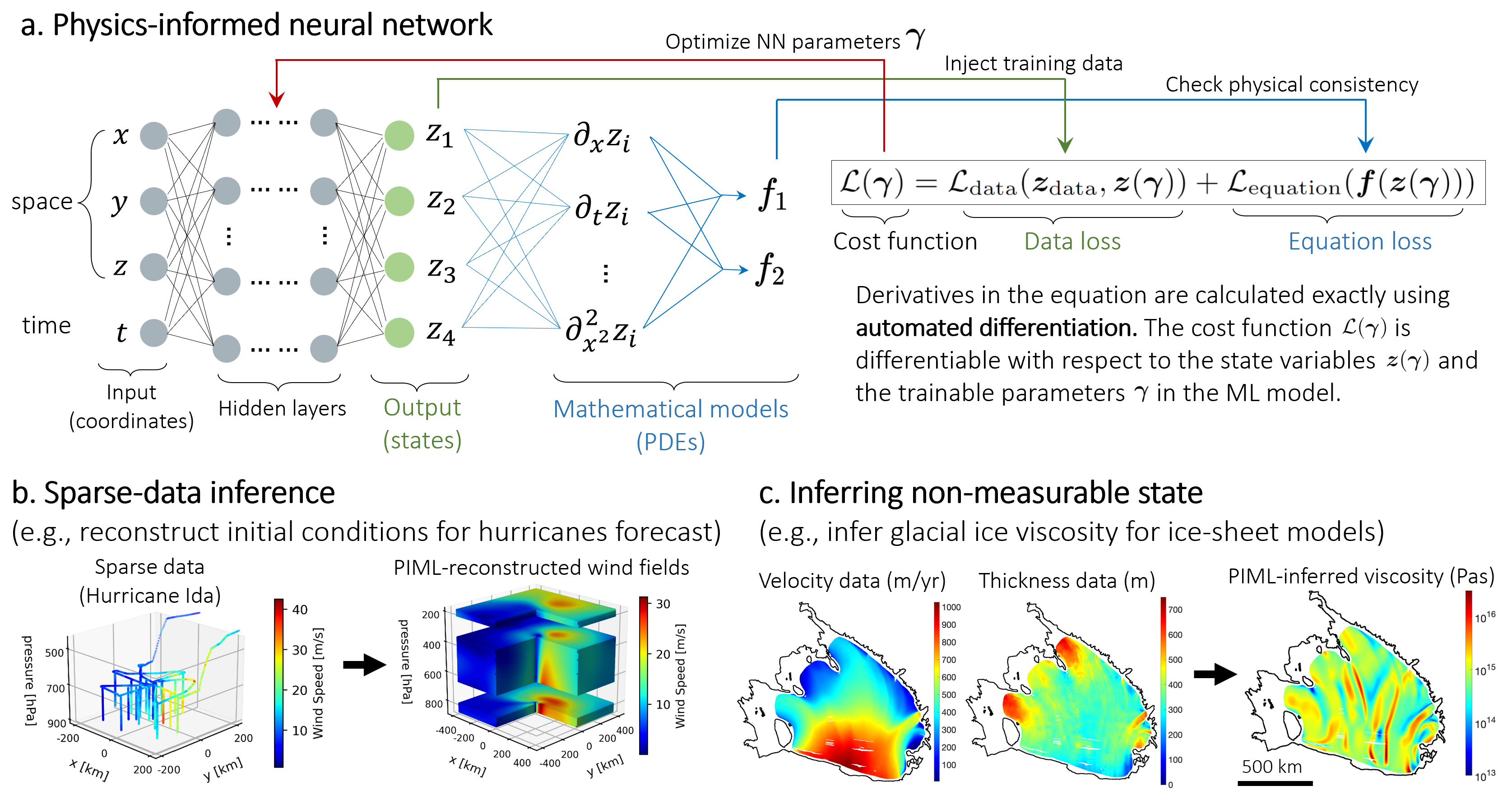} 
\caption{ (a) The physics-informed ML algorithm and its applications in data assimilation (b-c). Panels b and c adapted with permission from Eusebi et al. 2024 \cite{eusebi2024realistic} (CC BY 4.0) and Wang et al. 2022 \cite{wang2022discovering} (CC BY 4.0), respectively.}
\label{fig:PIML}
\end{figure}

\paragraph{Physics-informed neural networks (PINN)}\label{sec:pinn}
The use of PINNs \cite{raissi2019physics,karniadakis2021physics} for planetary-scale geophysical flow problems has started to emerge in the past few years. Introduced by Raissi et al. (2019) \cite{raissi2019physics}, PINN is a differentiable solver for partial differential equations (PDEs) that is particularly useful for inverse problems involving sparse-data inference, super-resolution, data denoising, and state estimation $\bm{z}$ in data assimilation (see sidebar titled Inverse problem and data assimilation). Unlike classical ML where the cost function typically only involves data, PINN encodes physics-based equations directly in the cost function (Fig. \ref{fig:PIML}a).

Throughout the training iterations, the optimizer identifies the best ML-parameterized states $\bm{z}=NN(\bm{x},t,\gamma)$ that are consistent with both the data and the governing equations. In the small-data regime, without evaluating the NN-parametrized $\bm{z}$ against known physical laws (such as conservation of mass, momentum, and energy), the ML predictions can be physically inconsistent and non-extrapolatable beyond the available observational data (e.g., deviation from truth in Fig. \ref{fig1}g). In contrast, by incorporating PDEs, PINN can achieve both physics-informed data interpolation and extrapolation, as demonstrated by the examples in Fig. \ref{fig:PIML}, which cannot be achieved by ML models trained with observational data alone.

Fig. \ref{fig:PIML}b-c demonstrate the applications of PINN on real observations, ranging from estimating the initial conditions $\bm{z}(\bm{x},t=0)$ of hurricanes for subsequent forecasts \cite{eusebi2024realistic} to inferring the non-measurable viscosity structure $z(\bm{x})$ of Antarctic ice shelves \cite{wang2022discovering}. Both examples fall within the ``small-data regime" in the upper left corner of Fig. \ref{datamodel}a, where incorporating knowledge of PDEs becomes crucial for solving the inverse problems; the PINN-reconstructed wind field $\bm{z}(\bm{x},t)$ (Fig. \ref{fig:PIML}b) involves only sparse observations of wind velocity itself as training data, obtained from measurements by hurricane hunter planes and dropsondes. The PINN prediction of ice viscosity is achieved without any observations of viscosity in the training data (Fig. \ref{fig:PIML}c); it relies solely on equations and other observable states (velocity and thickness fields) as training data. Thus, both examples involve substantial extrapolation beyond the sparse observational data, i.e., limited velocity data and no viscosity data. 

As long as the same data and physics-based equations are used to solve the inverse problem, the predictions generated by properly trained PINNs are as trustworthy as those produced by established data assimilation methods. Due to PINN's leverage of GPU and differentiable modeling to infer accurate initial conditions without ensembles of forward modeling, as used in ensemble-based data assimilation methods \cite{eusebi2024realistic} (see Supplemental Material for a brief comparison), PINNs require fewer computational resources to construct hurricane initial conditions with similar accuracy as the ensemble-based data assimilation methods \cite{lu2017gsi}. That being said, established data assimilation methods are supported by several mature theories, which are relatively lacking for PINN methods. Although several models used in climate predictions are not easily differentiable without substantial engineering efforts, the development of differentiable solvers for atmospheric dynamics \cite{kochkov2024neural} demonstrates promises. Differentiable ice-flow solver and emulator have also recently emerged as a new tool for forward and inverse modeling \cite{jouvet2023ice}.

\paragraph{Equation discovery} \label{eqn}
Existing equations $\bm{f}(\bm{z},\bm{\theta})$ describing the numerous processes in the climate system, particularly the SGS processes, are far from complete. Equation discovery, which outputs equations that are most consistent with data, has been used to tackle this problem.
Inspired by earlier symbolic regression algorithms for distilling physical laws from data \cite{schmidt2009distilling}, Sparse Identification of Nonlinear Dynamics (SINDy) \cite{brunton2016discovering} has emerged as a widely used method for discovering $\bm{f}(\bm{z},\bm{\theta})$ from data of the states $\bm{z}(t)$. It demonstrates the power of sparse regression for learning the most relevant terms in the prescribed function library that describes the data. To learn the correct $\bm{f}(\bm{z},\bm{\theta})$, SINDy requires sampled data of both $\bm{z}(t)$ and $d/dt({\bm{z}}(t))$. For many climate problems these state measurements are sparse and noisy, or entirely unavailable. 
Schneider et al. \cite{schneider2022ensemble} showed that time-averaged statistics of the states $\bm{z}(t)$, which are available for the climate system, can be sufficient to recover both the functional form of $\bm{f}(\bm{z},\bm{\theta})$ as well as the noise level of the data using sparse regression combined with EKI. Sparse EKI is robust to noisy data and was successfully implemented to recover the Lorenz 96 equations \cite{schneider2022ensemble}. Other approaches for equation discovery from data assimilation increments \cite{lang2016systematic,mojgani2023interpretable} and from partial observations \cite{chen2023causality}, motivated by climate problems, have been proposed too.

Arguably the most successful example of the application of equation discovery for climate physics so far has been the learning of an ocean mesoscale SGS \textit{parameterizations} \cite{zanna2020data}. Trained on high-resolution simulation data, Zanna and Bolton \cite{zanna2020data} showed that Bayesian linear sparse regression with relevance vector machines identifies relevant terms in the prescribed function library to discover the closed-form equations of $\Pi(\bar{z})$ (defined in Fig. \ref{fig1}) for eddy momentum and temperature forcing. The closed-form equation is consistent with an analytically derivable physics-based model \cite{anstey2017deformation, jakhar2023learning}. As discussed later in Section~\ref{SGS}, SGS parameterizations (Fig. \ref{fig1}) are essential for improving the accuracy of computationally feasible low-resolution climate simulations. While black-box NNs have also shown promises for developing data-driven SGS parameterizations (Section~\ref{SGS}), the significant interest in equation discovery stems from their better generalization to future climates and their interpretability (upper side of Fig. \ref{datamodel}b).

Inspired by early work on symbolic regression \cite{schmidt2009distilling,koza1994genetic}, the symbolic genetic algorithm \cite{chen2022symbolic} was developed to discover PDEs without the need to pre-determine a function library. It uses a binary tree to parameterize common mathematical operations (e.g., addition, multiplication, derivative, division) and finds the correct operations such that the discovered equation matches the data. In climate applications, genetic algorithms have been used for finding equations for cloud cover parameterization \cite{grundner2024data} and ocean parameterization \cite{ross2023benchmarking}.

\section{Machine Learning for Climate Simulations} \label{simulation}
We discuss two major directions leveraging ML methods to improve the accuracy of climate simulations: (a) SGS parameterization, aimed at developing more accurate climate models via better representation of small-scale (expensive to resolve) physical processes, and (b) Emulators, aimed at generating large ensembles of simulations (or directly, the statistics) at a fraction of the computational cost of a physics-based climate simulation. These two approaches are briefly discussed below.

\begin{textbox}[h]\section{Uncertainties in climate projections}
Climate projections are affected by three sources of uncertainty \cite{hawkins2009potential}: (a) \textit{model uncertainty} (also known as structural error), (b) \textit{internal variability uncertainty} (e.g., the signal-to-noise ratio problem), and (c) \textit{scenario uncertainty} (related to how much greenhouse gases will be released in the future). Reducing \textit{model uncertainty} requires developing more accurate climate models (e.g., improving parameterization or increasing resolutions), while reducing the \textit{internal variability} and \textit{scenario uncertainties} requires computationally efficient climate models that can generate long, large ensembles of simulations and explore different scenarios. 
\end{textbox}

\subsection{Subgrid-scale parameterization (SGS)} 
\label{SGS}
There are two reasons climate models require SGS parameterization to achieve simulations on relevant century-long timescales: (a) the process of interest varies on length or time scales smaller than a climate model's resolution, and (b) equations to describe the process are not known. SGS parameterizations estimate the effect of these unresolved processes on the resolved scales. Developing SGS parameterizations for climate modeling, but also for high-resolution simulations in limited domain, has been an active area of research since the pioneering work of Smagorinsky \cite{smagorinsky1963general} on the first climate models in the early 1960s. Still, the approximations made in formulating these parameterizations remain a leading cause of \textit{model uncertainty} (see the sidebar titled Uncertainties in climate projections). ML presents a potentially exciting path forward in improving these SGS parameterizations or developing new ones. The general idea is to use observations or high-resolution simulations to learn a data-driven representation or closed-form equation of the SGS term $\Pi$ (defined in Fig.~\ref{fig1}). Examples of the latter approach were discussed in Section~\ref{sec:algo}; below, we mainly focus on approaches based on NNs (Fig.~\ref{fig1}j).
\begin{marginnote}[]
\entry{Model uncertainty}{Deviation of the physics-based model from the data, could be due to inaccurate parameters $\theta$ or the physics equations $f$ itself.}
\entry{Structural error}{A type of model uncertainty arising from inaccuracies in the equations used to represent the data or processes of interest.}
\end{marginnote}

Studies have demonstrated the ability of ML algorithms such as NNs to learn parameterizations as a supervised learning task, $NN(\bar{\bm{z}}, \gamma) = \Pi(\bar{\bm{z}})$, for prototypes of geophysical turbulence (e.g., \cite{zanna2020data, guan2022stable}), ocean turbulence (e.g., \cite{bolton2019applications,sane2023parameterizing}), moist convection and clouds (e.g., \cite{gentine2018could,gentine2021deep,yuval2023neural,rasp2018deep,grundner2022deep,arcomano2023hybrid, watt2024neural}), and atmospheric gravity waves (e.g., \cite{matsuoka2020application,espinosa2022machine,hardiman2023machine}). Some of these examples have achieved stable simulations that are more accurate than simulations with traditional physics-based parameterizations (e.g., \cite{rasp2018deep,zanna2020data,guan2022stable,arcomano2023hybrid}). However, while promising, this approach faces a number of challenges. Some are common among other ML applications to climate (e.g., interpretability, extrapolation) and are discussed in Section~\ref{sec:challenges}. Challenges specific to SGS parameterization via supervised learning include availability of suitable high-resolution training data from numerical studies or observational campaigns, and issues with accuracy and stability once these ML-based SGS parameterizations are coupled (e.g., with atmosphere \cite{rasp2018deep,pahlavan2024explainable} and ocean models \cite{zanna2020data}).

This discussion of SGS parameterization for ``climate models" would not be complete if we did not emphasize the distinct challenges compared to its use in ``weather forecasting". In weather forecasting, the objective is to predict a specific trajectory based on initial conditions, necessitating accurate and detailed prediction of SGS physics. Conversely, climate studies aim to predict changes in the system's average behavior over decades. Thus, it is sufficient to predict the SGS statistics rather than all its specific features. 
This shift requires novel ML approaches optimized to capture emergent statistics rather than detailed information from training data in supervised learning. This poses challenges as long term observations of climate statistics are limited and the simulations coupled with ML-based SGS parameterizations need to be stable for long enough timescale to learn the climate statistics from the training data. Despite these challenges, a few studies have made progress in producing stable and accurate simulations of simple climate prototypes using NNs trained with differentiable modeling \cite{frezat2022posteriori} and EKI \cite{pahlavan2024explainable} that target the evolution of the climate variables in response to the SGS processes rather than training on the SGS processes themselves. Recently, Google Research has made strides in this direction by developing an atmospheric model's dynamical core that learns SGS physics statistics directly from reanalysis data~\cite{kochkov2024neural}. Yet, numerical stability and satisfaction of global energy conservation constraints remain a challenge. Unlike in the weather literature, training from direct observations \cite{mcnally2024data} has not been attempted yet possibly because of the sparsity of global datasets with long enough timescale to capture climate statistics.

An additional challenge is how to address the interactions between different SGS processes, for example, between boundary layer turbulence in the ocean and atmosphere. Training is typically done on sub-components of the full climate system because it is not computationally feasible to run global climate simulations that fully resolve all SGS processes and their interactions. Concurrent observations of different SGS processes are also limited. As a result, interactions among a number of individually trained/calibrated data-driven parameterizations can lead to inaccurate or even unstable global simulations. This is an area in need of practical advancements.

\begin{textbox}[h]\section{Reanalysis}
Reanalysis, sometimes referred to as ``maps without gaps", refers to a method of using a physical model to assimilate disparate observational data streams into a combined multivariate dataset uniform in space and time. The model fills in the data-poor regions and ensures physical consistency between variables. The output is often referred to as \emph{reanalysis data} but it is important to keep in mind that these data are not observations, but outputs of a forecast model. In fact reanalysis products from different weather centers will differ amongst themselves, and this spread can be taken as a measure of uncertainty in observations and understanding.
\end{textbox}

\subsection{Climate emulators} \label{emulation} 
``Emulator'' refers to several types of tools in the climate science literature. In general, an emulator is trained to mimic the data, from physics-based simulations or observations, to substantially reduce the computational cost of producing new climate predictions, e.g., for other climate conditions within the distribution of the training data.

Emulators can be used to interpolate the projections from expensive climate simulations, making their projections among different emission scenarios accessible without re-running the simulations. Earlier use of ML for emulators followed the successful approach of traditional pattern-scaling emulators \cite{beusch2020emulating,tebaldi2022stitches}, which, for example, predict the change in statistics of variables of interest (e.g., regional annual-mean surface temperature or the return period of extreme events at a later time) given a small set of inputs (e.g., year, greenhouse gas forcing, global mean surface temperature). Using ML techniques (e.g., GPR, NN), emulators such as ClimateBench~\cite{watson2022climatebench} have been employed to estimate the climate impacts of anthropogenic emissions annually up to 2100. However, it remains to be demonstrated that their skill is superior to that of pattern-scaling emulators, i.e. emulators that regress regional temperature on global mean temperature or cumulative emissions.

\paragraph{Spatio-temporal emulators}
While the aforementioned emulators can predict aggregated statistics within an often large window of length and time scales, another type of emulator has emerged in recent years with the aim of predicting the evolution of the climate system at fine spatio-temporal scales. These ``spatio-temporal'' emulators leverage the success of ML-based weather forecast models, which are physics-free and trained solely on reanalysis data \cite{hersbach2020era5} (spanning 1979-present; see sidebar titled Reanalysis). Recent ML-based weather forecast models (e.g., FourCastNet \cite{pathak2022fourcastnet}, Pangu \cite{bi2023accurate}, GraphCast \cite{lam2023learning}) are time-stepping algorithms that solve the \textit{initial-value problem} of predicting the state $\bm{z}(t)$ of the global atmosphere forward in time (from $t_i$ to $t_{i+1}$, then from $t_{i+1}$ to $t_{i+2}$, and so on; Fig. \ref{fig1}i). They exhibit comparable or even better skill than the best physics-based weather prediction models for lead times of up to around 10 days~\cite{lam2023learning}. However, weather and climate predictions are different problems. The former is an \textit{initial value problem} while the latter is more akin to a \textit{boundary value problem} in the sense that the focus is on how external boundary conditions impact the system over longer periods of time. 
\begin{marginnote}[]
\entry{Spatio-temporal emulator}{An emulator with predictions that evolve with space and time, e.g. weather forecasting, sea ice seasonal forecasting.}
\entry{Initial value problem}{A forward model $\bm{f}(\bm{z})$ with predictions $\bm{z}(t)$ subject to the initial conditions $\bm{z}(t_0)$ specified at time $t_0$.}
\entry{Boundary value problem}{A forward model $\bm{f}(\bm{z})$ with predictions $\bm{z}(t)$ subject to the boundary conditions $\bm{z}(x_0,t)$, that can be time-evolving, at the boundaries $x_0$.}
\end{marginnote}

For climate predictions, atmospheric spatio-temporal emulators are built to solve \textit{boundary-value problems} that integrate the global atmospheric state given external forcings (e.g. radiative forcing) and time-evolving boundary conditions (e.g., sea-surface and land temperature) for decades or centuries. The AI2 Climate Emulator (ACE) \cite{watt2023ace,duncan2024application} is a promising example of such a spatio-temporal emulator trained on physics-based simulations. Similar work on oceanic spatio-temporal emulators \cite{bire2023ocean,subel2024building} suggests that {\it coupled} climate emulators might start to emerge as well. 

ML spatio-temporal emulators have shown even more promise in simulating components of the climate system whose physics are less well understood. For the cryosphere, deep learning-based emulators for seasonal sea-ice prediction have been found to outperform state-of-the-art physics-based dynamical models in terms of forecast accuracy \cite{andersson2021seasonal,wang2023subseasonal,zhu2023deep}, with a lead time of a few months. Some of these sea-ice emulators capture atmospheric-ice-ocean interactions by training with appropriate climate variables \cite{zhu2023deep,andersson2021seasonal}. Because these emulators were trained directly on sea ice observational data, they learn the atmospheric-ice-ocean interactions that are incompletely parameterized in the physics-based dynamical models, thereby correcting the model's \textit{structural error}).  

\begin{textbox}[h]\section{Challenges and opportunities of ML-based emulators}
Emulators can address the climate response to a particular emission \textit{scenario} and \textit{internal variability} uncertainty. However, emulators are at best as accurate as the data they are trained on, which may still contain \textit{model uncertainty}; this may be partially overcome by training with data from very high-resolution yet very expensive simulations, such as the emerging global 1-km runs. Nonetheless, major questions about the stability and physical consistency of the trained spatio-temporal emulators need to be addressed. For example, ML weather forecast models have been shown to produce unstable or unphysical atmospheric circulations beyond 10 days, poorly represent small scales processes~\cite{chattopadhyay2023long,ben2024rise}, and do not reproduce the chaotic behavior of weather \cite{selz2023can}. Potential solutions to address these challenges include incorporating physical constraints into ML models (Section \ref{PIML}) and developing a deeper understanding of the different sources of error in these models (Section \ref{UQ}-\ref{sec:OOD}). \end{textbox}

\subsection{Physics-informed machine learning (PIML)} \label{PIML} 
Despite ML's ability to emulate weather (Section \ref{emulation}) and parameterize SGS processes when trained on high-resolution simulations or observations (Section \ref{SGS}), there is no guarantee that its predictions are physically sound (e.g., conserve mass, energy). This physical inconsistency is problematic and makes long-term climate projections using ML-based emulators and ML-based parameterizations not trustworthy (Fig. \ref{fig1}g, left panel).
Incorporating physics constraints such as conservation laws, symmetries, and more broadly, equivariances (defined below), has been shown to alleviate a number of challenges such as instabilities and learning in the small-data regime--Kashinath et al.~\cite{kashinath2021physics} review earlier work in physics-informed ML for weather and climate modeling.
\paragraph{Conservation laws} Various methods exist for incorporating conservation laws into ML models, such as embedding them in the loss function (e.g., physics-informed NNs \cite{raissi2019physics}; Section \ref{sec:pinn}) or other components of the ML architecture. For instance, Beucler et al. \cite{beucler2021enforcing} demonstrated that conserving quantities like mass and energy can be enforced as hard constraints within the NN architecture. Their architecture-constrained NN, trained as a SGS parameterization of moist convection, significantly improved simulated climate.
\paragraph{Symmetries and equivariances} Incorporating symmetries and equivariances has also shown advantages, particularly in the small-data regime. For a variable $x$, the nonlinear function $g$ is equivariant under transformation $\mathbf{A}$ if $\mathbf{A}g(x)=g(\mathbf{A}x)$. For example, by incorporating various symmetries (e.g., scale equivariance, rotational equivariance) into convolutional neural networks (CNNs) trained on turbulence data from previous time steps, the CNNs generalize well to future time steps \cite{wang2020incorporating} (Fig. \ref{fig1}h). Enforcing rotational equivariance through Capsule NN, CNNs, or customized latent spaces has improved ML-based predictions of large-scale weather patterns \cite{chattopadhyay2022towards} and turbulent flows \cite{guan2023learning, wang2020incorporating}.
\paragraph{Spectrum information}
Including information about the Fourier spectrum of geophysical turbulence in the loss function has been shown to aid in learning small scales and reducing \textit{spectral bias}, thereby improving the stability and physical consistency of ML-based emulators \cite{chattopadhyay2023long} (see Section \ref{UQ} for further discussions).

\section{Challenges and Promises} \label{sec:challenges}
\subsection{Quantifying uncertainties of ML models} \label{UQ}
Broadly speaking, there are two sources of error in ML-based models: errors in the training \textit{data} and the \textit{epistemic uncertainty} for the ML model. The errors in data can stem from sparsity and measurement noise, which are particularly relevant for observations, or from errors in simulation data, which can arise from numerical errors and inaccurate physics-based equations. The \textit{epistemic uncertainty} of the ML model arises from different sources, such as model architecture and hyperparameters \cite{psaros2023uncertainty}. Some ML techniques, such as GPR, provide rigorous estimates of uncertainty (e.g., \cite{watson2021machine, watson2022climatebench} for climate applications). However, for deep learning, uncertainty quantification (UQ) is more complicated and the subject of extensive research (see recent review papers in the context of scientific applications \cite{psaros2023uncertainty,abdar2021review}). 
\begin{marginnote}[]
\entry{Epistemic uncertainty}{Deviation of the ML model from the data, could be due to approximation, optimization, and generalization errors.}
\end{marginnote}

Understanding the sources of errors in ML models can improve their stability, physical consistency, and reliability. For example, in simulations coupled with ML-based parameterization, errors from the ML model are propagated into the simulations and vice versa, potentially leading to instabilities and nonphysical behavior. Similarly, errors in a spatio-temporal emulator can accumulate and destabilize the emulation. Since we cannot directly estimate "accuracy" during inference (as we do not have access to the ground truth), the best approach is to estimate the "uncertainty" of the ML model's output, as this uncertainty may be indicative of its accuracy. Here, we provide examples from climate science for uncertainty quantification of NNs, and we also discuss two impactful sources of \textit{epistemic uncertainties} related to representation error (e.g., spectral bias) and data imbalance (e.g., rare extreme events).

\paragraph{Quantifying epistemic uncertainty}
A variety of techniques from the ML literature have been employed for UQ of NNs in climate applications. For instance, deep ensembles \cite{sonnewald2021,Yik2024, mansfield_uncertainty_2024,clare2021} and Bayesian NNs \cite{clare2021,sun2024data} are used to assess the mean and spread of predictions, and the faithfullness of the NN optimization. In \cite{draeger2024}, these two methodologies were combined to reveal the consequences of architecture choice, as determined by UQ and the ability to approximate the physical system. Other techniques, such as variational autoencoders, dropout, and abstension, have also been explored \cite{guillaumin2021stochastic,foster2021probabilistic,sun2024data,barnes2021controlled}. See \cite{haynes2023creating,sun2024data} for detailed discussions.



\begin{figure}
    \centering
    \includegraphics[width=0.75\linewidth]{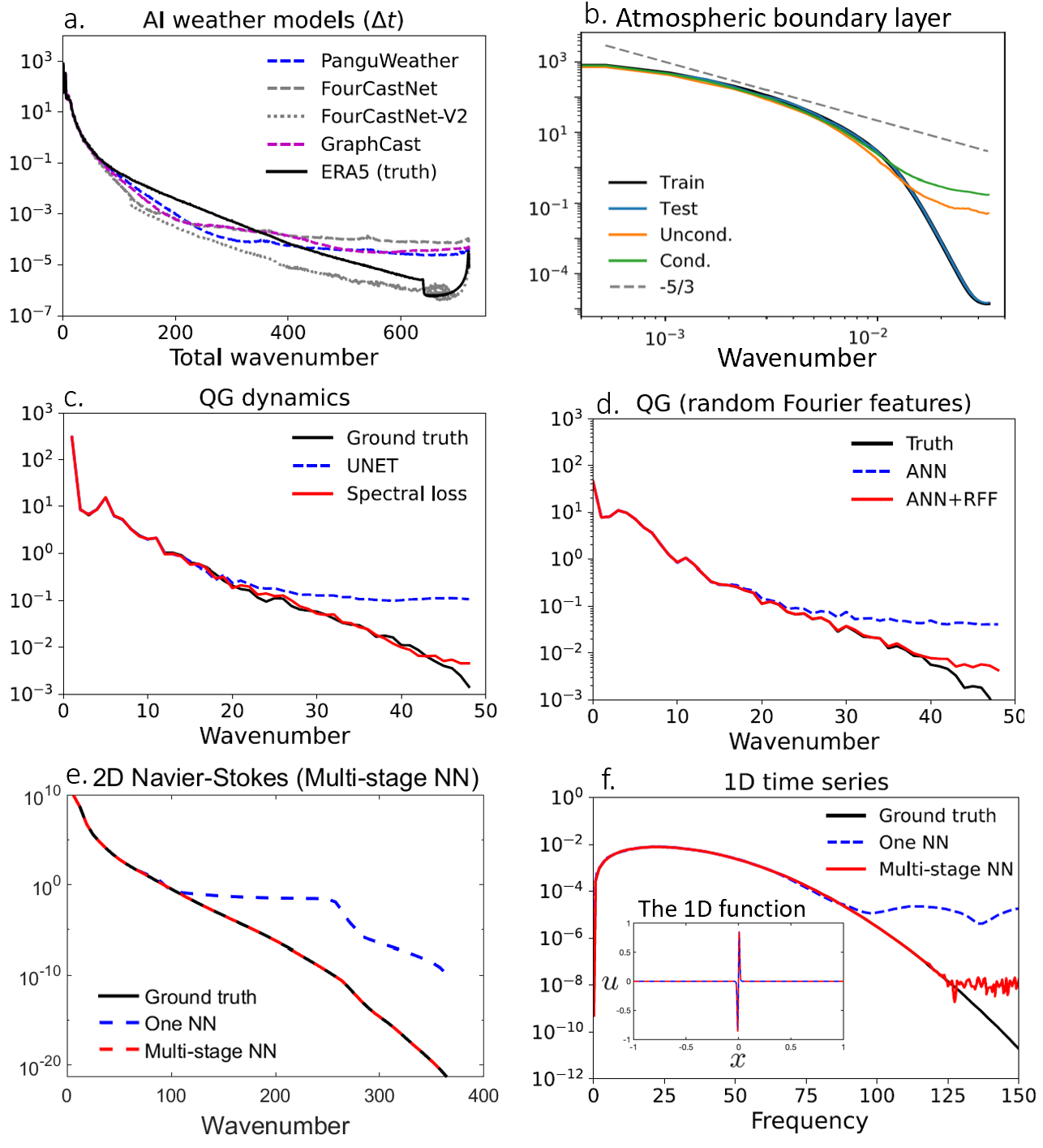}
    \caption{The spectral bias of neural networks \cite{rahaman2019spectral, xu2019frequency} can be widely observed in climate applications and can cause major challenges such as instabilities. (a) State-of-the-art ML-based weather emulator predictions after one time step (based on results from \cite{chattopadhyay2023long}, courtesy of Qiang Sun). (b) Atmospheric boundary layer turbulence reconstruction \cite{rybchuk2023ensemble}. (c-d) Quasi-geostrophic (QG) turbulence prediction after one time step \cite{chattopadhyay2023long} (courtesy of Ashesh Chattopadhyay) or reconstruction (based on results from \cite{mojgani2023interpretable}, courtesy of Rambod Mojgani). (e) Reconstruction of a 2D flow field \cite{ng2024spectrum} using the multi-stage NN \cite{wang2024multi}. (f) Reconstruction of a 1D function with sharp peaks, difficult to fit with vanilla NNs (Courtesy of Yongji Wang). Panels a, b, c, d, e adapted with permission from Chattopadhyay and Hassanzadeh 2023 \cite{chattopadhyay2023long} (CC BY 4.0), Rybchuk et al. 2023 \cite{rybchuk2023ensemble} (CC BY 4.0), Mojgani et al. 2024 \cite{mojgani2023interpretable} (CC BY 4.0), Ng et al. 2024 \cite{ng2024spectrum} (CC BY 4.0), respectively.
    }
    \label{fig:spectral}
\end{figure}

\paragraph{Spectral bias}
Another example of \textit{epistemic} error in ML models affecting climate applications is the ``\textit{spectral bias}" \cite{rahaman2019spectral} (or frequency principle \cite{xu2019frequency}). Namely, NNs learn to represent the large scales much more easily than small scales, which can pose challenges for multi-scale climate problems. Fig. \ref{fig:spectral}a shows the spectrum of the one-time-step ($\sim$ 6 or 12h) prediction of upper-level wind from a few state-of-the-art ML weather emulators. While the predictions exhibit the correct spectrum for up to zonal Fourier wave numbers of $\sim 30$ (scales of 40000~km to 500~km), smaller scales (from 500~km to 25~km) are poorly learned. It's noteworthy that these predictions all boast around $\sim 99\%$ accuracy based on anomaly pattern correlation. These errors in small scales grow to larger scales after 10 days. Eventually, these predictions either blow up or become unphysical \cite{chattopadhyay2023long}. The same behavior is observed in simpler tasks such as reconstruction of atmospheric boundary layer turbulence (Fig. \ref{fig:spectral}b), time-stepping prediction for quasi-geostrophic (QG) turbulence (Fig. \ref{fig:spectral}c) and its re-construction (Fig. \ref{fig:spectral}d), or even a simple 1D function (Fig. \ref{fig:spectral}f). Promising solutions include Fourier regularization of the loss function (Fig. \ref{fig:spectral}c from \cite{chattopadhyay2023long}) and random Fourier features \cite{tancik2020fourier,wang2021eigenvector} (Fig. \ref{fig:spectral}d from \cite{mojgani2023interpretable}). Superposing small NNs \cite{wang2024multi, ng2024spectrum} via the multi-stage NNs also improves spectral bias substantially compared with vanilla NN (Fig. \ref{fig:spectral}e). 
\begin{marginnote}[]
\entry{Spectral bias}{NNs’ tendency to preferentially capture certain frequencies of the training data.}
\end{marginnote}

\paragraph{Rare extreme events} 
Another example of \textit{epistemic} error relates to rare events (e.g., heat waves, hurricanes, ice-shelf collapse, ocean circulation collapse). Predicting these rare events is crucial, but they are often underrepresented or entirely absent from the training set, leading to significant ``data imbalance''. Addressing data imbalance and improving the learning of rare events in an active area of research. Common approaches such as re-sampling \cite{miloshevich2023probabilistic,lopez2023global}, using weighted loss function \cite{sun2024data,rudy2023output}, and learning the causal relationship that drives the rare behavior \cite{draeger2024} have shown promise. Innovative approaches, such as combining ML-based emulators with mathematical tools for rare events \cite{ragone2018computation,finkel2021learning}, may enable the learning of the rarest events.  

\subsection{Non-stationarity: out-of-distribution error of ML models} \label{sec:OOD} 
Climate change is inherently non-stationary \cite{palmer1999nonlinear}: the mean state and its variability change over time. This poses a major challenge for applications of ML models to climate-change projections. For instance, ML-based models trained on data from the current climate may not perform well for a warmer future climate with higher greenhouse gas concentrations. Studies have already demonstrated unstable or unphysical simulations resulting from NN's inability to extrapolate beyond its training data \cite{rasp2018deep,subel2023explaining,beucler2024climate}. The non-stationary problem raises new questions: How can we ensure that the prediction task is within the distribution of the training data? How do we ensure that the ML model leverages information that is climate-invariant? Would a hybrid approach, coupling physics- and ML-based models, improve long-term climate simulations?

Examples of strategies for dealing with non-stationarity and out-of-distribution generalization include: (a) incorporating physical knowledge \cite{beucler2024climate} and (b) transfer learning \cite{subel2023explaining, guan2022stable}. As an example of (a), the recently proposed ``\textit{climate-invariant}" ML \cite{beucler2024climate} learns the mapping between variables of interests that is universal across climates. This study showed promising offline results of data-driven parameterization of moist convection across a range of cold to warm climates once temperature, relative humidity, and latent heating were properly transformed. This approach leverages physical insights of climate. The main challenge is finding the appropriate transformations, which are easier to find for thermodynamics but harder to find for dynamics (e.g., wind).
\begin{marginnote}[]
\entry{Climate-invariant ML}{ML that utilizes relationships that stay the same across different climates to improve generalization.}
\end{marginnote}

Transfer learning, a common framework in ML for addressing out-of-distribution generalization, involves training an ML model for a given system (e.g., the current climate) and then re-training it with a much smaller amount of data from a new system (e.g., a warmer climate). The re-trained ML model could then perform better for the new system. Several studies have demonstrated the potential of transfer learning to address significant changes in parameters $\theta$ (e.g., a 100 times increase in Reynolds number in geophysical turbulence) or forcing \cite{subel2023explaining, guan2022stable}. The key challenge with transfer learning is obtaining reliable data for re-training. In climate change prediction, we must rely on simulations, as observations from the future are unavailable. Libraries of high-resolution global and regional simulations that strategically sample from a range of climates are emerging, providing a valuable source of training and re-training data \cite{shen2022library,sun2023quantifying,satoh2019global}.


\begin{figure}
    \centering
    \includegraphics[width=1\linewidth]{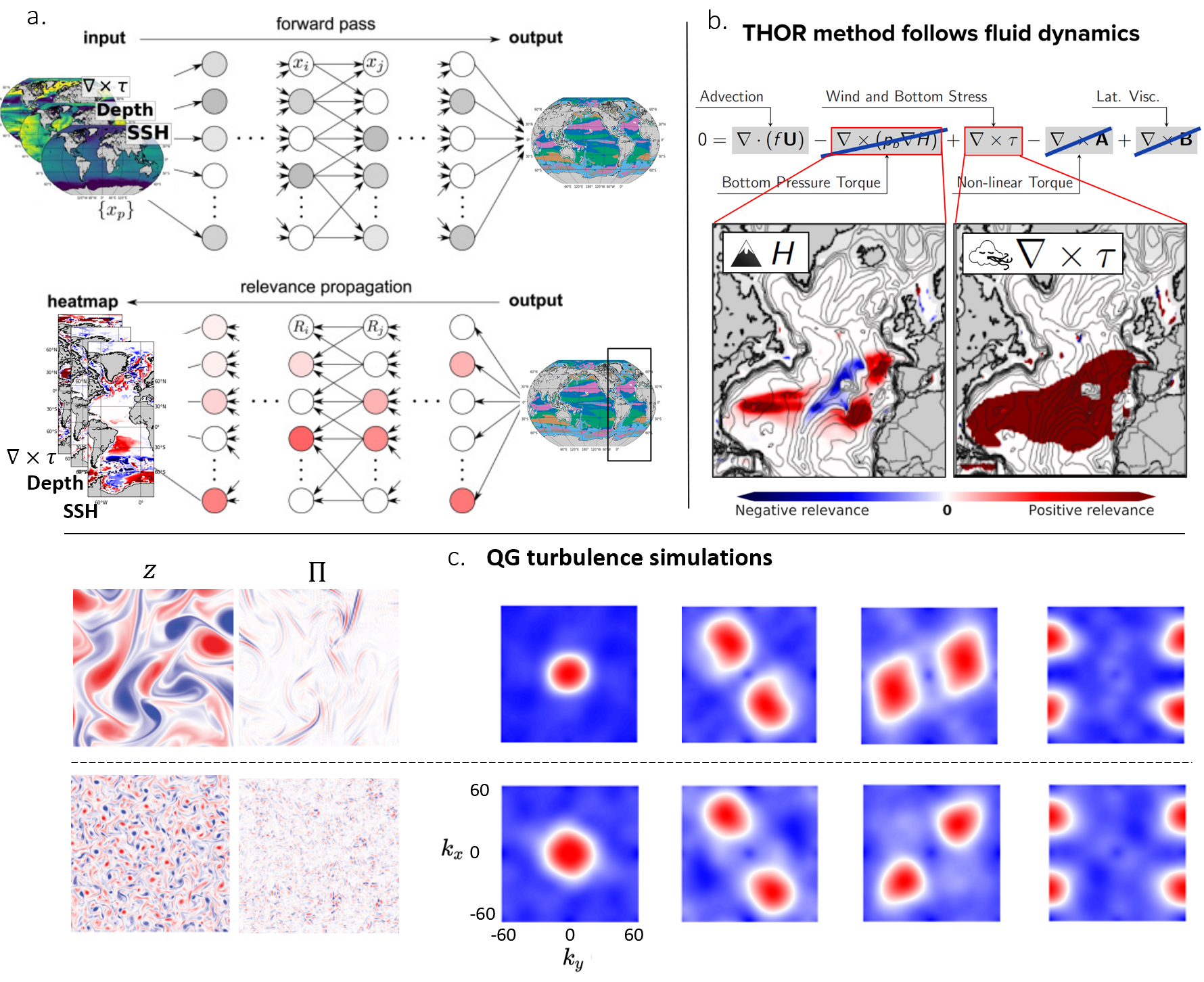}
    \caption{Understanding what ML learns. Panel (a, b) is a cartoon illustration of ensuring that the input data necessary for the ML model to demonstrate its learning of physics is present from the THOR method (Tracking global Heating with Ocean Regimes) \cite{sonnewald2021}. (a, top) Training a NN to predict sections of the ocean dominated by different balances in equation terms describing the flow (colors in output) using related surface fields (e.g., wind). (a, bottom) Looking \textit{backwards} using XAI to see where in the input the NN saw as relevant (blue: not relevant, red: relevant). (b) For the pink section in the North Atlantic, only two equation terms are relevant (red boxes), and relevances show conformance in two maps below, e.g., where the mountain range (close black lines) in depth (H) gives negative relevance. (c) The two leftmost panels show examples of the state $\bm{z}$ and SGS term $\Pi$ (defined in Fig. \ref{fig1}) from two setups of geophysical turbulence that differ in forcing scale and dynamics. The other panels show examples of the Fourier spectra of convolutional kernels of NNs trained as ML-based SGS parameterizations $NN(\bar{\bm{z}},\gamma)=\Pi$. The Fourier analysis shows the emergence of low-pass, high-pass, and band-pass Gabor filters \cite{subel2023explaining}. Panels a-b and c adapted with permission from Sonnewald and Lguensat 2021 \cite{sonnewald2021} (CC BY 4.0) and Subel et al. 2023 \cite{subel2023explaining} (CC BY 4.0), respectively.}
    \label{fig:XAI}
\end{figure}

\subsection{Understanding what ML is learning}\label{XAI}
Understanding what an ML model has learned and how and why is essential for climate applications, especially to gain trust and further improve such models. The use of Explainable AI (XAI) techniques from the ML community for climate applications has gained popularity in recent years (e.g., \cite{sonnewald2021, Mamalakis2022, camps2020advancing,mayer2021subseasonal}). For review papers featuring XAI in climate science, see \cite{bommer2024finding, Flora2024,Irrgang2021, Sonnewald2021_review}. A core strategy for many XAI techniques is to identify which parts of the inputs of an NN (e.g., regions of the atmosphere or ocean) are used to predict a specific output. For example, Toms et al.~\cite{toms2020physically} illustrated XAI could infer scientifically meaningful information regarding climate patterns known as El Nino events. In Labe and Barnes \cite{Labe2022}, XAI was used to assist model comparisons of the Arctic. XAI is also often used to determine whether the ML has gained skill through detecting meaningful patterns in the training data, instead of spurious correlations. Sonnewald et al. \cite{sonnewald2021} used XAI within an ensemble of NNs to determine their ML model's accuracy (Fig. \ref{fig:XAI} top panels a and b) by assessing conformance with theory. The ML model's task was to predict ocean physical `regimes,' i.e., dominant balances between terms in the equation governing the flow. Similar equation-determining frameworks are in \cite{draeger2024, Yik2024, clare2021}.

Standard techniques for analyzing physical systems can also be applied to understand NNs. For example, Fourier analysis has provided insight into NNs' learning process \cite{rahaman2019spectral,xu2019frequency}. In the context of data-driven modeling of geophysical turbulence \cite{subel2023explaining}, Fourier analyses of CNNs revealed what they have learned. The convolution kernels (with over 1 million learnable parameters) were shown to fall into just a few classes: low- and high-pass filters, and Gabor wavelets (Fig. \ref{fig:XAI}c). These findings align well with prior work on using wavelets for turbulence modeling \cite{farge1992wavelet}, and even more so with theoretical ML studies on the need for such spectral filters for learning multi-scale, localized data \cite{mallat2016understanding,olshausen1996emergence}. More recent work has found this approach useful in interpreting deep NNs in climate applications by examining concepts from physics and ML together \cite{pahlavan2024explainable, chattopadhyay2023long}.

\section{Summary and Outlook} 
Numerous scientific discoveries and rigorous understandings have been driven by empirical relationships. We summarized in Section \ref{discovery} how ML can facilitate this, for example, by accelerating the search for patterns in climate data that can be used to derive physical understandings. We also summarized in Section \ref{DIMD} the promises ML has brought to find closed-form equations for poorly understood climate processes. In many aspects of the climate system, we do not yet have accurate process-level models to describe the system (e.g., sea ice rheology and cloud microphysics). The increasing amount of observational data offers exciting opportunities for both equation and knowledge discovery to improve the fundamental understanding of climate physics.

On the other hand, ML can be used as tools to improve simulations. ML models can be coupled with traditional physics-based models and used to parameterize processes for which closed-form equations are not yet available (Section \ref{SGS}). ML has led to breakthroughs in weather forecasting, a task not widely expected to be possible a couple of years ago. We discussed the challenges to overcome when moving forward from weather forecasting to climate prediction (Sections \ref{emulation} and \ref{sec:challenges}). 

ML is advancing rapidly, and new techniques and concepts that have shown great promise in other fields are now being quickly adopted in climate science. Notable examples, as of this writing, include diffusion models (e.g., \cite{price2023gencast,bassetti2023diffesm,finn2024towards}), large language models (e.g., \cite{zhou2024proof}), and foundation models (see, e.g., \cite{mukkavilli2023ai} for a discussion of their design and implementation and \cite{gupta2024machine} for a downstream task involving gravity waves). Progress in climate modeling could greatly benefit from collaborations between the ML, climate sciences, and mathematics communities. For example, the numerical analysis of differential equations and the advent of digital computers played a key role in starting the field of numerical weather and climate prediction \cite{balaji2021climbing}. Developing similar rigorous tools, by closely combining methods from climate physics, ML theory, and numerical analysis, can potentially help with building stable, accurate, and trustworthy ML-based models.

\section*{DISCLOSURE STATEMENT}
The authors are not aware of any affiliations, memberships, funding, or financial holdings that might be perceived as affecting the objectivity of this review. 

\section*{ACKNOWLEDGMENTS}
We thank Mingjing Tong, Oliver Dunbar, Jinlong Wu, and Duncan Watson-Parris for their helpful discussions regarding data assimilation methods, GPR with EKI, sparse learning, and emulators, respectively. We are grateful for the valuable general feedback from Andre Souza and Janni Yuval on this article. We also thank Qiang Sun, Ashesh Chattopadhyay, Rambod Mojgani, and Yongji Wang for helping to remake the spectral bias figure. C.Y.L., R.F., P.H., and A.S. acknowledge the National Science Foundation for funding via Grants No. DMS-2245228, AGS-2426087, OAC-2005123, and OAC-2004492, respectively. R.F., P.H. and A.S. also acknowledge funding from Schmidt Sciences through the Virtual Earth System Research Institute.

\bibliographystyle{ieeetr}
\bibliography{reference}

\end{document}